\newcommand{\p}{\partial}
\newcommand{\f}[2]{\frac{#1}{#2}}
\newcommand{\der}[2]{\frac{\rm\,d#1}{\rm\,d#2}}

\documentclass{mn2e}
\usepackage{epsfig}

\begin{document}
 
\title[Conservative Mass Transfer.]{The Different Fates of a Low-Mass X-ray Binary. I: Conservative Mass
Transfer}

\author[G. Lavagetto et al.]{G. Lavagetto,$^1$\thanks{email: lavaget@gifco.fisica.unipa.it} L. Burderi,$^2$ 
F. D'Antona,$^2$ T. Di Salvo,$^3$, 
R. Iaria$^1$ and N. R. Robba$^1$ \\
$^1$ Dipartimento di Scienze Fisiche ed Astronomiche, 
Universit\`a di Palermo, via Archirafi n.36, 90123 Palermo, Italy.\\
$^2$ Osservatorio Astronomico di Roma, Via Frascati 33, 
00040 Monteporzio Catone (Roma), Italy.\\
$^3$ Astronomical Institute "Anton Pannekoek", University of 
Amsterdam and Center for High-Energy Astrophysics,\\
 Kruislaan 403, NL 1098 SJ Amsterdam, the Netherlands.}
\maketitle
\begin{abstract}
 We study the evolution of a low mass x-ray binary coupling a binary 
stellar evolution code with a general relativistic code that describes the
behavior of the neutron star.
We assume the neutron star to be low--magnetized ($B \sim 10^8$ G).
 In the systems investigated in this paper, our computations show that 
during the binary evolution
the companion transfers as much as $1~M_{\odot}$ to the neutron star, with 
an accretion rate of $\sim 10^{-9}~M_{\odot}/{\rm yr}$. This is sufficient 
to keep the inner rim of the accretion disc in contact with the neutron star 
surface, thus preventing the onset of a propeller phase capable of 
ejecting a significant fraction of the matter transferred by the companion.
In this scenario we find that, for neutron stars governed by equations of 
state from soft up to moderately stiff, an accretion induced collapse to a 
black hole is almost unavoidable.
The collapse to a black hole can occur either during the accretion phase 
or after the end of the mass transfer when the neutron star is left in a 
supramassive sequence. In this last case the collapse is driven by energy
losses of the fast spinning magneto-dipole rotator (pulsar). For extremely 
supramassive neutron stars these energy losses cause 
a spin up. As a consequence the pulsar will have a much 
shorter lifetime than that of a canonical, spinning down radio pulsar.
This complex behavior strongly depends on the equation of state  for ultra-dense 
matter and therefore could be used to constrain the internal structure of the 
neutron star.
In the hypothesis that the r-modes of the neutron star are excited during the accretion process,
 the gravitational waves emisson limits the maximum spin attainable
by a NS to roughly 2 ms. In this case, if the mass transfer is conservative, the collapse to a black hole
during the accretion phase is even more common since the maximum mass achievable 
before the collapse to a black hole during accretion is smaller due to the limited spin frequency. 
\end{abstract}
\section{Introduction}

The widely accepted scenario for the formation of millisecond radio pulsars
(MSP) is the  recycling of an old  neutron star (hereafter NS) by a spin-up 
process. The spin-up is due to the accretion of matter and angular momentum 
from a Keplerian disc that is fueled {\it via} Roche lobe overflow of a binary 
late-type companion (see Bhattacharya \& van den Heuvel 1991 for a review). 
If enough mass and angular momentum are transferred, the NS 
spin attains an equilibrium value that is roughly equal to
the keplerian angular frequency at the inner rim of the accretion disc 
(Ghosh \& Lamb, 1979). Since the NS has a weak surface magnetic field 
($\sim 10^8$ G), the magnetospheric radius (at which the disc pressure is 
balanced by the magnetic pressure) truncates the accretion disc close by or 
at the NS surface, and the equilibrium period is expected to be, in most 
cases, below one millisecond.
Typically $\sim 0.35\; M_\odot$ are sufficient to reach a spin 
period of 1 ms (e.g.\ Burderi et al. 1999). Most donor stars in systems hosting
recycled MSPs have certainly lost, during their interacting binary evolution, 
a mass greater than $0.35\; M_\odot$ since they now appear as white 
dwarfs of mass $0.15-0.30\; M_\odot$ (e.g.\ Taam, King, \& Ritter 2000), whose 
progenitors are likely to have been stars of $1.0-2.0\; M_{\odot}$ (Webbink, 
Rappaport, \& Savonije 1983; Burderi, King, \& Wynn 1996; Tauris \& Savonije 
1999). Therefore \textit{if the mass transfer is conservative} 
the amount of matter accreted is well sufficient to spin the 
star up to periods below 1 ms 
(Cook, Shapiro, \& Teukolsky 1994a), or even to produce an accretion induced
collapse into a black hole. Once the accretion  
and spin-up  process ends, the magnetospheric radius  expands beyond the light 
cylinder radius (where an object corotating with the NS attains the speed of 
light). This initiates a phase in which the rotational energy of the NS is emitted 
{\it via} electromagnetic radiation and the star can be observable as a 
rapidly rotating radio pulsar. According to this model, Low Mass X-ray Binaries 
(hereafter LMXBs) are the progenitors of MSPs. Indeed the discovery of
coherent X-ray pulsations in four transient LMXBs, namely  SAX J1808.4--3658 
with a spin period $P=2.5~\mathrm{ms}$ and an orbital period of 
$P_{\mathrm{orb}} = 2~\mathrm{h}$ (Wijnands \& van der Klis 1998), 
XTE J1751--305 ($P=2.3~\mathrm{ms}$, $P_{\mathrm{orb}} = 42~\mathrm{min}$, 
Markwardt et al. 2002), XTE J0929--314 ($P=5.4~\mathrm{ms}$, 
$P_{\mathrm{orb}} = 43~\mathrm{min}$, Galloway et al. 2002), and XTE J1807--294 
with a spin period $P=5.2~\mathrm{ms}$ and an orbital period of 
$P_{\mathrm{orb}} = 40~\mathrm{min}$ (Markwardt et al. 2003) confirmed that 
NSs in LMXBs can be accelerated to millisecond periods. 

Several numerical methods have been developed to solve the Einstein equations 
for a rotating NS (see Stergioulas 1998 for a review). Stability criteria 
show that a rapidly rotating NS can support a maximum mass (against 
gravitational collapse) much higher than the non-rotating mass limit, since the 
centrifugal force attenuates the effects of the gravitational pull 
(e.g.\ Friedman, Ipser, \& Sorkin 1988).
Conversely, if a rotating NS has a mass that exceeds the non-rotating limit 
(i.e. a supramassive NS), it will be subject to gravitational collapse if 
it loses enough rotational energy. Numerical simulations of rotating  NSs  
show that, contrarily to the standard behavior, supramassive NSs spin up just 
before collapse, even if they lose energy. This effect is known to be stronger
for higher mass objects (Cook, Shapiro, \& Teukolsky 1992).

The value of the maximum rotating and non-rotating mass depends on the equation 
of state (EOS) governing the NS matter.  On the other hand, the minimum allowed 
period for a given mass occurs when gravity is balanced by centrifugal 
forces at the NS equator (mass shedding limit). Thus the spin period can 
be used to constrain the mass-radius relation for the NS, i.e. its EOS. 
In the context of the standard (gravitationally-bound) NS models 
(e.g.\ Glendenning 2000), several EOSs have been proposed. 
We usually distinguish different EOSs depending on their 
stiffness (i.e. the value of ${\rm d} p /{\rm d} \epsilon$, were $p$ is the 
fluid pressure and $\epsilon$ is the energy density). If the EOS is stiff, the 
 matter is less compressible at high densities, resulting 
in larger NS radii as compared to a soft EOS, and hence in longer minimum 
rotational periods. Except for few, very stiff cases, most EOS predict minimum 
rotational periods well below 1 ms. However, no sub-millisecond pulsars have 
been detected up to date: the shortest observed spin period is  
$\sim 1.5~ \mathrm{ms}$ (Backer et al. 1982), uncomfortably higher than 
the theoretical predictions. 

In an attempt to find an explanation for the apparent clustering of the spin periods of millisecond pulsars
around 2 ms, Bildsten (1998) and Andersson (1998) indipendently suggested that
LMXBs emit gravitational waves once they reach a critical spin frequency.
Burderi and D'Amico (1997) showed that for nonaxisymmetric  m-modes,
 assuming a realistic range of temperatures, the values of the critical spin frequency 
are remarkably close to the limiting spin frequency determined by the centrifugal limit at the border of the NS. 
On the other hand, Andersson, Kokkotas and Stergioulas (1999)
demonstrated that at a certain spin frequency (much lower than the maximum attainable spin period)
an instability to the Rossby waves (r-modes) of the star arises, thus causing emission of gravitational waves.
Levin (1999) suggested that the gravitational waves emission causes the onset of a spin-up spin-down cycle,
and not of  a steady state spin equilibrium: in this scenario the NS  undergoes a very rapid spindown
 (lasting $\sim 1$ yr) due to the rapid heating during the r-mode excitation, and then 
starts another cycle of accretion driven spin up. Brown and Ushomirsky (2000) showed that
if the NS has a superfluid core the steady state scenario is not viable because the
predicted quiescence luminosity in this case is much higher than the observed one in the X-Ray transient Aql X-1.
 Levin and Ushomirsky (2001) showed that, when keeping into account the presence of the solid crust, the critical spin frequency for 
the onset of the r-mode varies between $\sim  600$ Hz and $\sim 200$ Hz, depending on the core temperature.

In this paper we consider the full evolution of a LMXB,  trying to 
determine how the results of our modeling of the recycling scenario compare 
to the observations and which effects peculiar to General Relativity are 
indeed observable. 
We will also show the differences  in the evolution of the system when the r-mode instability is excited during 
accretion and when it is not excited.

\section{Evolution equations for the compact object}

A rotating NS is unambiguously defined by the
boundary conditions for the integration of the Einstein equations or, 
equivalently, by suitable pairs of resulting integrated  quantities, such as 
the baryonic mass and the angular momentum or the baryonic mass and the 
total mass--energy of the star. Therefore the evolution of the NS is determined by 
the evolution of such pairs. We consider the evolution of the NS both 
during the accretion phase and after the accretion has finished: in the former 
case it is convenient to solve the evolution equations for baryonic mass and 
angular momentum, since we know the general formula for the torque exterted by the accreting matter on the NS,
 while in the latter we solve the evolution equations for 
baryonic mass and mass--energy, since we can evaluate the luminosity of a magnetodipole rotator using
Larmor's formula.

\subsection{Evolution equations for the accretion phase}
According to accretion theories, matter transferred from the companion star 
to the NS via Roche-Lobe overflow forms an accretion disc around the compact 
object. The disc is truncated because of one of the following reasons:
i) the interaction with the magnetic field
of the NS, which truncates the disc at the magnetospheric radius $r_m$; 
ii) the presence of NS surface itself at $R_{\rm NS}$; iii) the lack of 
closed Keplerian orbits for 
radii smaller than the marginally stable orbit radius, $R_{\rm MSO}$ 
(at few -- depending on the mass and spin of the 
compact object -- Schwarzschild radii from the NS centre).

The magnetospheric radius is defined as the radius at which the 
pressure of the disc equals the pressure of the magnetic field of the NS.
The magnetospheric radius can be written as a fraction $ \phi$ (see Burderi 
et al. 1998) of the Alfv\'en radius $R_{\rm A}$ (the radius at which the energy
density of the, assumed dipolar, NS magnetic field equals the kinetic
energy density of the spherically accreting matter) as 
\begin{eqnarray}
\nonumber
r_{\rm m}&=& \phi \; R_{\rm A}=2.45 \times \alpha^{9/35} 
m_G^{1/28} \dot{m}_B^{3/70} r_{\rm m}^{-3/28} R_{\rm A}\\
&=&2.244 \times \left(\alpha^{36/5} m_G \dot{m}_B^{6/5}  R^{28}_{\rm A} \right)^{1/31} \,\mathrm{cm}
\label{eq:rm}
\end{eqnarray}
where $\alpha \le 1$ is the Shakura-Sunyaev parameterization of the 
accretion-disc viscosity (for which we will assume a typical value of $0.1$), 
 $m_G$ is the NS
gravitational mass in $M_{\odot}$,  $\dot{m}_B$ is the baryonic accretion rate 
on to the NS in $10^{-8}~M_{\odot}/{\rm yr}$, and $R_{\rm A}$ is
\begin{equation}
R_{\rm A} = 1.24 \times 10^6 
\mu_{26}^{4/7} m_G^{-1/7} \dot{m}_B^{-2/7} \;{\rm cm}
\end{equation}
where $\mu_{26}$ is the magnetic dipole moment 
of the NS in units of $10^{26}~{\rm G\; cm^3}$ , defined from $B_s=\mu/R^3$, where R is the 
equatorial radius and $B_s$ the surface magnetic field of the NS at its equator.

The NS radius, that is always in the order of $10^6 ~\mathrm{cm}$, 
depends  both on the mass of the NS and on its angular momentum. In general 
smaller radii correspond to larger masses while larger radii correspond to 
larger angular momenta. Thus a rapidly rotating NS can have a much
larger radius than a non-rotating one (the equatorial radius can expand 
up to $40\%$, see Cook, Shapiro, \& Teukolsky 1994b).

\begin{figure*}
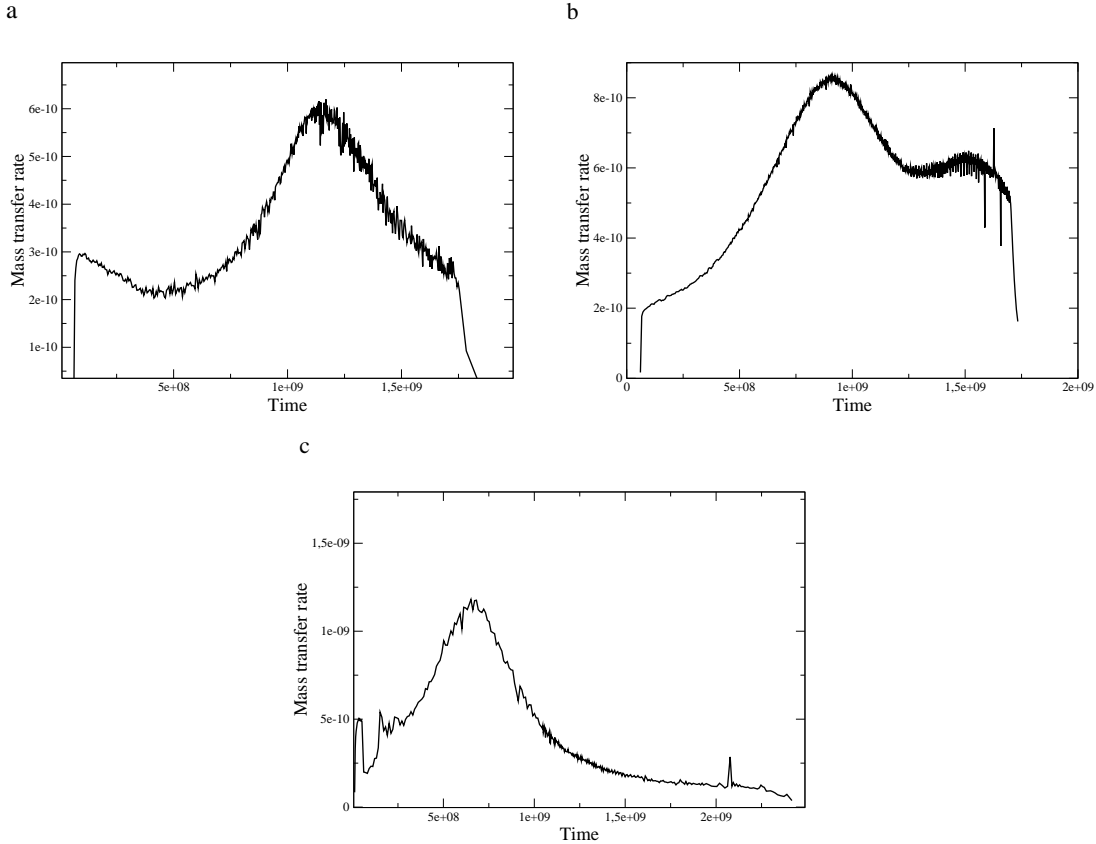

  \centering
$$ \epsfig{figure=fig1a.eps}\qquad \epsfig{figure=fig1b.eps}$$
$$ \epsfig{figure=fig1c.eps}$$

  \caption{ Baryonic mass accretion rates (in $M_{\odot}/yr$) as a function of time (in yr) for
systems consisting of a NS with initial gravitational 
mass of $1.4 M_{\odot}$ and: a) a population II (low metallicity) 
star of $0.85 M_{\odot}$  transferring about $0.65 M_{\odot}$;
 b)  a $1.15 M_{\odot}$ population I star transferring about $0.91 M_{\odot}$; 
c) a $1.199 M_{\odot}$ population I star transferring about $0.99 M_{\odot}$.}
\label{fig1}
\end{figure*}

The marginally stable orbit is the smallest stable orbit possible for a test 
particle around an axisymmetric, rotating body of gravitational mass $M_G$ and 
angular momentum J. Its radius, following 
Bardeen (1970), is:
\begin{eqnarray*}
  R_{\mathrm{MSO}}&=& R_{\rm g} (3+Z_2- \sqrt{(3-Z_1)(3+Z_1+ 2Z_2)});\\
  Z_2&=&\sqrt{3 \left(\f{a}{R_{\rm g}} \right)^2+Z_1^2}\\
  Z_1&=&1 + \left(1+ \left(\f{a}{R_{\rm g}} \right)^2 \right)^{1/3}+ 
\left(1- \f{a}{R_{\rm g}}\right)^{1/3}\\
  a&=& \f{J}{M_G c}\\
  R_{\rm g}&=& \f{GM_G}{c^2}.
\end{eqnarray*}
where $c$ is the speed of light and $G$ is the gravitational constant.
Therefore the inner radius of the disc, $R_D$, will be:
\begin{equation}
  \label{eq:rd}
R_D = \max \left\{r_\mathrm{m}, R_{\rm NS}, R_{\rm MSO} \right\}.   
\end{equation}

\begin{figure}
  \centering
  \epsfig{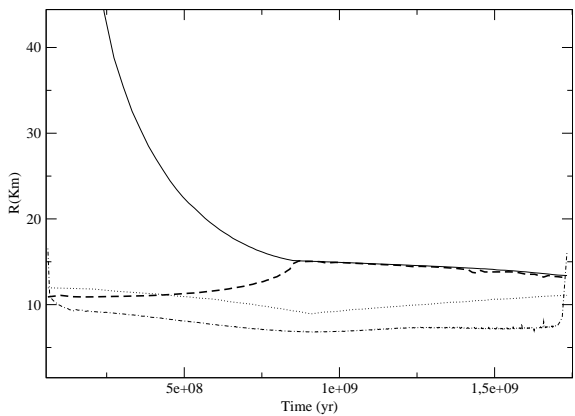}
  \caption{ Time evolution of the relevant radii for the
accretion process on to the neutron star in the binary system of figure 1b.
The EOS governing the ultradense matter is EOS FPS. 
The solid line is the corotation radius, the thick dashed line is NS radius,
the dot dashed line is the magnetospheric radius, the dotted line is the 
radius of the marginally stable orbit. Note that the magnetospheric radius 
exceeds the corotation radius only at the end of accretion, when
the mass transfer rate decreases, and that for the first part of the accretion process
the last stable orbit is outside the NS. The magnetospheric radius data are 
smoothed in order to avoid the disturbing visual effects due to the 
numerical instabilities of $\dot{M}_B$.}
  \label{fig:radii}
\end{figure}

The position of the inner rim of the disc is crucial for the
accretion of matter on to the NS. In particular it is very important in the case 
$R_D=r_{\mathrm{m}}$:
since at $r_\mathrm{m}$ the matter is forced by the magnetic field to corotate 
with the NS, accretion on to a spinning magnetized 
NS is centrifugally inhibited once $r_{ \mathrm{m}}$ lies outside 
the corotation radius $r_{\rm co}$, the radius at which the Keplerian 
angular frequency of the orbiting matter is equal to the NS angular frequency:
 $ r_{\rm co} = 1.50 \times 10^{6} m_G^{1/3} P_{-3}^{2/3}$ cm,  
where $P_{-3}$ is the spin period in milliseconds.    
Conversely if the magnetospheric radius is smaller than the corotation radius
 accretion of matter can proceed undisturbed.
However, as $r_m$ scales as a negative power of $\dot{m}_B$, a decrease in 
the mass transfer rate, which can occur for instance at the end of the 
accretion phase, will results in an expansion of the magnetosphere.
 In this paper we considered NSs with  
$\mu_{26} \simeq 1$, which is typical of observed MSPs (see e.g.\ 
Lorimer 1994), corresponding to surface magnetic fields of $\sim 10^8~\mathrm{G}$.
Moreover, in the cases considered here, the accretion rate is always above 
$\sim 10^{-10} M_{\odot}/\mathrm{yr}$
during the mass transfer phase (see figure~\ref{fig1}). In this case, 
eq.~(\ref{eq:rm}) implies a magnetospheric radius 
of $\sim 10^6$ cm, that is comparable with the NS radius.  
Although we considered in our simulations (see below) the possibility 
of a propeller phase, in these conditions the details of the interaction 
of the accretion disc matter with the NS magnetosphere during the accretion 
phase will not affect much our results. In fact the magnetospheric radius 
lies outside of the corotation radius only for a small time ($\la 0.1 \%$ of 
the evolution time) just before of the end of accretion (see 
figure~\ref{fig:radii} for details). In other words,
the spin evolution of a low magnetized NS is not very different from the 
spin evolution of a non--magnetized NS, in which no propeller effect is 
possible since the disc is always truncated near the surface of the NS.
We can therefore write the mass evolution equation for $r_m < r_{co}$, which is 
simply given by the condition that the baryonic mass per unit time accreted 
on to the NS is equal to the mass lost by the companion per unit time due 
to Roche-Lobe overflow, $\dot{M}_{RL}$, i.e.

\begin{equation}
  \label{eq:mbdot}
  \dot{M}_{B} = \dot{M}_{RL}.
\end{equation}

Lamb, Pethick, \& Pines (1973) wrote a general equation describing the flow 
of angular momentum into the stellar magnetosphere including the material 
stress at the inner edge of the disc as well as magnetic and viscous stresses 
in a disc partially threaded by the NS magnetic field. 
In particular they assumed that the threading occurs in a transition region 
near the inner rim of the disc  where the magnetic field of the NS couples to 
the accretion disc. However they noticed that, for slow rotators, magnetic and 
viscous stresses can be neglected with respect to the contribution of the 
material stress. Ghosh, Lamb, \& Pethick (1977) showed that, in the 
case of a disc rotating in the same sense of the star, if the transition 
region where the magnetic field couples to the disc is small, these additional 
contributions are still negligible even for rapidly rotating NS.
In the present discussion we assume therefore that the torque exerted on the 
NS is only due to the contribution of the material stress at the inner edge of 
the disc (Pringle \& Rees 1972). We will study the evolution of high-magnetized 
NSs with heavily threaded accretion discs in a future paper. 
Note, however, that Wang (1996) showed that threading effects can modify the 
maximum achievable period (i.e.\ the \textit{equilibrium period}) of the NS
only by a few percents with respect to the unthreaded case.

 The angular momentum per unit baryonic mass of a particle orbiting around a 
rotating, axisymmetric object at a distance R is (see Bardeen 1970) 
\begin{equation}
  \label{eq:elle}
 j (R) = \f{ (GM_G)^{\f{1}{2}} (R^2 - 2 (J/M_G c) (R GM_G/c^2)^{\f{1}{2}} 
+ (J/M_G c)^2)}{R^{\f{3}{4}} (R^{\f{3}{2}} -3 (GM_G/c^2) R^{\f{1}{2}} +
 2 (J/M_Gc)(GM_G/c^2)^{\f{1}{2}})^{\f{1}{2}}} .
\end{equation}
Thus we can write the equation for angular momentum evolution of the NS 
simply as
\begin{equation}
  \label{eq:j}
\left\{\begin{array}{c}
\dot{J}= j(R_D) \cdot \dot{M}_B, \hskip 1cm J < J_{max} (M_B) \\
\dot{J}= \der{J_{max}}{M_B}  \dot{M}_B, \hskip 1.2 cm J = J_{max} (M_B)
\end{array}
\right.
\end{equation}
The first equation does not apply when the NS is at the mass shedding, i.e. the 
regime in which the matter at the border of the star has the Keplerian velocity
at that radius; for each value of baryonic mass, the mass shedding regime is 
individuated by the corresponding maximum angular momentum of the star, 
$J_{max} (M_B)$. At mass shedding, the matter of the disc should dissipate 
angular momentum to accrete on to the star.
If the disc matter has an angular momentum that will make the star exceed 
the mass shedding limit, it will just not be able to accrete since it  will 
not be gravitationally bound to the star, until viscous forces drive the 
excess of angular momentum to the outer zones of the disc (where it will be 
given back to the companion through tidal forces) allowing matter to accrete on to the star. 
In this situation, the second equation holds. Integrating it we get
$J_f- J_i= J_{\mathrm{MAX}, f} - J_{\mathrm{MAX}, i}$, but since $J_i = J_{\mathrm{MAX}, i}$ we have
$J_f=J_{\mathrm{MAX}, f}$:
the accreted matter will give the NS only the angular 
momentum that keeps it at mass shedding, and the star will continue to move 
along the maximum rotation line.

\subsection{Evolution equations at the end of the accretion phase}
The NS is thought to switch on as a radio pulsar when the inner edge 
of the disc lies outside the light cylinder radius (i.e. the radius 
at which a particle corotating with the star will have velocity $c$), 
$r_{lc}= c / \omega_{NS}$ - where $ \omega_{NS}$ is the angular velocity of the NS:
this certainly happens when the accretion stops and thus the disc disappears.
 The emission mechanisms for a radio pulsar are believed to be rotating magnetic dipole radiation 
and magnetospheric currents associated with the emission of relativistic particles, 
both depending on the angle i between the NS magnetic moment $\mu$ and the spin axis (Goldreich \& Julian 1969). 
These two emission mechanisms compensate in such a way that the total energy emitted is nearly independent 
of i (Bhattacharya \& van den Heuvel 1991). The energy loss per unit time will be
$\dot{E}= - \f{2}{3 c^3} \mu^2 \omega_{NS}^4.$
 It is then convenient to describe the NS evolution in terms of 
baryonic mass and total energy rather than in terms of baryonic mass and 
angular momentum as we did for the accretion phase. 
We can write the  evolution equations as
\begin{eqnarray}
  \label{eq:evpul}
\nonumber
  \dot{M}_{B}& =& 0\\
  \dot{M}_G&=& - \f{2}{3 c^5} \mu^2 \omega_{NS}^4.
\end{eqnarray}
Depending on the different conditions, we use either equations (\ref{eq:mbdot}) 
and (\ref{eq:j}) or equations (\ref{eq:evpul}) to compute the spin evolution of 
the NS.
\subsection{Excitation of r-modes}

Since our main goal is to compare qualitatively the evolution of systems in which 
the gravitational waves damping is not present and systems in which it is present, 
we will not go into the details of the theory of r-modes excitation. We will simply 
assume that, if the r-mode mechanism is present, any star attaining a period of 2 ms
(corresponding to a frequency of 500 Hz) undergoes a rapid spindown to a period of 
5 ms due to gravitational waves emission following the scenario proposed by Levin (1999).
 After this rapid spin-down, the NS starts again its accretion driven spin up process.

\subsection{Numerical methods of integration}

 During accretion phases we coupled through equations (\ref{eq:mbdot}) and 
(\ref{eq:j}) a detailed description of the binary evolution of the system 
obtained with the stellar evolution code  
with a fully relativistic calculation of NS
physical properties given its EOS, its baryonic mass and its angular 
momentum.

The evolution of the binary system is followed selfconsistently including the
full computation of the structure of the secondary star, by means of the
ATON1.2 stellar evolution code (D'Antona, Mazzitelli \& Ritter, 1989). The
equations of stellar structure are numerically solved by a full
Newton--Raphson integration from the center up to the basis of the stellar
atmosphere. The numerical inputs are described in Mazzitelli (1989). The
secondary star is considered to be the mass losing component of the binary
system, and its mass loss rate is computed following the formulation  by
Ritter (1988), as an explicit exponential function  of the distance of the
stellar radius to the Roche lobe, in units of the pressure scale height.
The evolution of the binary parameters can be followed by considering
several possible cases for the transfer of mass, and for the loss of mass and
associated angular momentum from the system. The orbital evolution
also includes losses of orbital angular momentum through magnetic braking, in
the Verbunt and Zwaan (1981) formulation, in which the braking parameter is
set to $f=1$\  and through gravitational radiation.

The relativistic computations for the compact object are implemented
using a slightly modified version of RNS (Rotating Neutron Stars) 
public domain code by Stergioulas \& Morsink (1999).
 The RNS routines provide 
the numerical solution of Einstein equations for a rotating axisymmetric body, 
integrating the equations \textit{via} Komatsu-Eriguchi-Hachisu method 
(Komatsu et al. 1989; see Stergioulas \& Friedman 1995 for a comparison with 
other integration methods). The main problem of this approach is that 
we have evolution equations for baryonic mass and angular momentum of the 
compact object, while the boundary conditions for the solution of Einstein 
equations are the central energy density 
and the oblateness of the star.   To solve this problem, we used a grid of 
relativistic equilibrium NS models integrating the Einstein equations for a 
wide range of initial conditions, spanning all allowed values for stable 
configurations. This was obtained modifying the code to build a complete grid 
of relativistic equilibrium configurations with the necessary numerical 
precision. To this we added a stability control routine to exclude unstable 
configurations.

 We obtain the accretion rate $\dot{M}_B$ and a corresponding time interval 
(in which $\dot{M}_B$ remains unchanged) from the stellar evolution code. 
Then we integrate the differential equations (\ref{eq:mbdot}) and (\ref{eq:j}) 
using a finite differences method. For each integration time 
step the accreted baryonic mass is $\Delta M_B = \dot{M}_B (t) \Delta t$
and the accreted angular momentum is 
$\Delta J= j(R_D(t)) \dot{M}_B (t) \Delta t$. 
For the $n$-th time step of the evolution, we search on the grid the single 
equilibrium configuration with the updated values of 
$M_B(t_n)= M_B(t_{n-1}) + \Delta M_B $ and $J(t_n)=J(t_{n-1})+ \Delta J $;
in this way we get the corresponding values of gravitational mass, radius, 
spin frequency and momentum of inertia of the NS.
We fine-tuned the time step  so that $\Delta M_B$, $\Delta J$ are always
comparable with the distance of two neighbour points in the grid. 

 Obviously the integration over a grid may introduce numerical uncertainties 
in our results: in particular, we are implicitly assuming that the spin 
frequency, the gravitational mass and the radius of the NS remain unchanged 
between two points 
on the grid. Therefore we introduce an error in the evaluation of these 
quantities that is equal to half the distance between two points on the 
grid, that is always well below $0.5\%$ in our simulations. 
This also yields an uncertainty in evaluating, step by step, the value of the 
accretion radius given by eq.~(\ref{eq:rd}) and the angular momentum of the 
matter at the inner rim of the disc, that is always in the order of a few 
percent.

During the pulsar phase, we integrated equations~(\ref{eq:evpul}) in a similar way 
using the same grid of relativistic equilibrium NS models. To avoid the grid 
imprecisions we made a cubic spline interpolation on the values of the 
function $\omega_{NS} (M_G)$ for the integration of the equation set (\ref{eq:evpul}).

\section{Results}

We consider three simple examples that show different 
possible fates for a LMXB, depending on the characteristics 
of the companion star. In all cases we start our evolution 
with a slowly rotating ($P \sim 1~{\rm s}$) NS with a gravitational mass of 
1.4 $M_{\odot}$ -- which is a typical value for a newborn NS (Thorsett and Chackrabarty, 1999). The
EOS adopted to describe the ultradense matter is that proposed by 
Friedman, Pandharipande and Smith (FPS, see e.g. Lorenz et al. 1993 for a recent discussion)
which has been widely used in the literature and has average values
of stiffness and physical parameters when compared to other EOS 
(see Arnett and Bowers, 1977 for a catalog). For our simulations we consider 
the maximum mass configuration equal to the maximum rotation configuration 
(although this is not precisely
true, see Stergioulas \& Friedman 1995, these configurations differ one from each other for less than $0.1 \%$ in our case),
 and we have 
$M_{\mathrm{max}}= 2.123~M_{\odot}$, $\omega_{\mathrm{max}}/2\pi=1882~{\rm Hz}$.
Finally, as already mentioned, we assume that the NS has a magnetic dipole 
moment of $10^{26}\; {\rm G~cm}^3$.

In particular we consider three typical examples of binary evolution in 
which the accretion rate is low enough to remain conservative, both above and 
below the bifurcation period -- i.e.\ the period below which the orbital 
evolution, during the accretion phase, proceeds towards shorter binary periods.
The bifurcation period is $P_{\rm orb,bif} \simeq 18$ hr for a binary
composed of a $1.4\; M_\odot$ NS plus a $1.0\; M_\odot$
secondary (Podsiadlowski et al. 2002). In the cases we considered, the 
secondary is:
\begin{enumerate}
\item  A population II donor  with an 
initial mass of $0.85 M_{\odot}$ and an orbital period above the orbital bifurcation period;
 it loses $0.646 M_{\odot}$ during the 
accretion phase. The initial binary period is $14.3$ hours and it steadily 
increases to about $124$ h at the end of the accretion phase, which lasts  
$\sim 1.7 \times 10^{9}$ yr (see figure 1a), leaving a helium white dwarf 
remnant. Note that the orbital bifurcation period 
for a system composed of a  $1.4\; M_\odot$ NS plus a $0.85 M_{\odot}$ is smaller 
than $14.3$ hours, because the bifurcation period depends strongly on the companion mass.
In this case, Roche-Lobe overflow is driven by the nuclear evolution 
of the companion.  
\item  A population I donor (initially below the bifurcation period) with 
an initial mass of $1.15 M_{\odot}$, which loses $0.91 M_{\odot}$ during the 
accretion phase. The orbital period evolves from $10.6$ to $3.5$ hours. In this case the 
companion star overflows its Roche-Lobe due to angular momentum losses
caused magnetic braking. 
The accretion phase ends when the companion becomes fully convective, according to 
the classical scenario for the evolution of cataclysmic variables in the 
period gap (Verbunt \& Zwaan 1981). The accretion phase lasts 
$\sim 1.7 \times 10^{9} {\rm yr}$, see figure 1b.
\item A population I donor (initially above the bifurcation period) with an initial 
mass of $1.199 M_{\odot}$; it loses $0.99 M_{\odot}$ during the accretion 
phase. The period evolves from $19$ to $92$ hours, leaving again a helium 
white dwarf remnant. The accretion phase lasts 
$\sim 2.2 \times 10^{9} {\rm yr}$ (see figure 1c).  In this case, Roche-Lobe 
overflow is driven by the nuclear evolution of the secondary star.
\end{enumerate}
\begin{figure*}
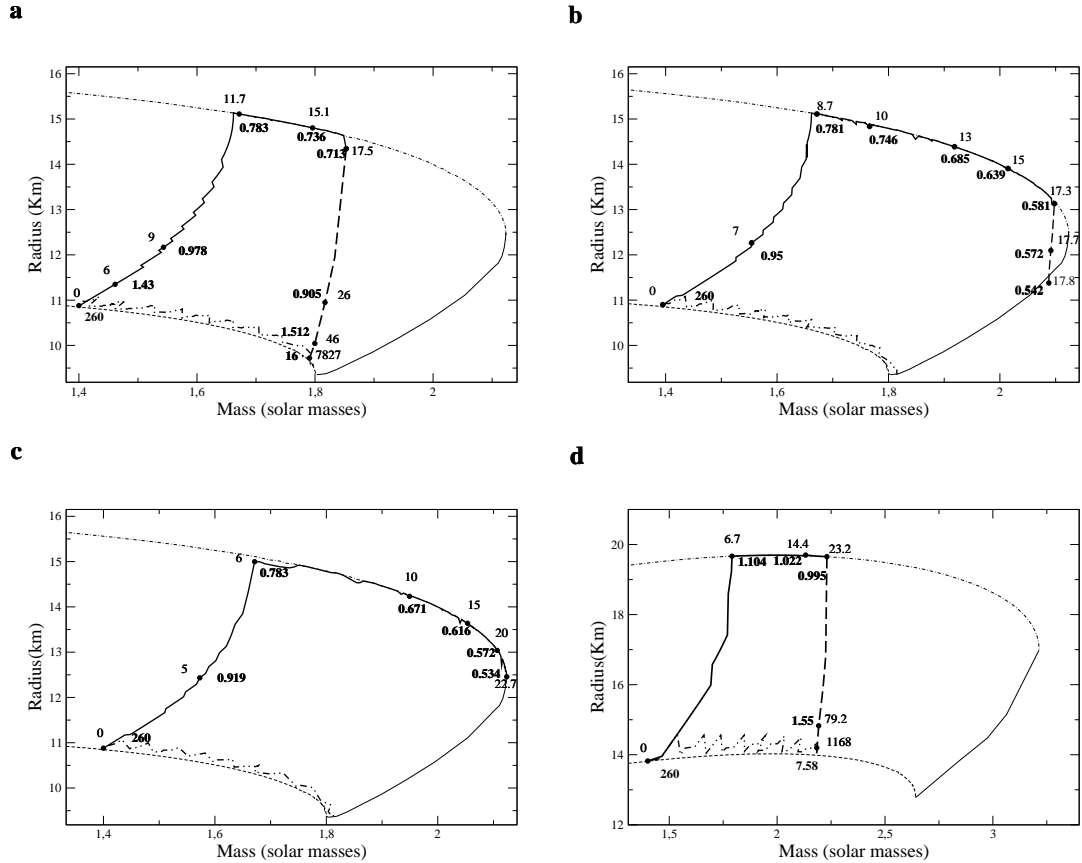

  \centering
$$ \epsfig{figure=fig3a.eps}\qquad \epsfig{figure=fig3b.eps}$$
$$ \epsfig{figure=fig3c.eps}\qquad \epsfig{figure=fig3d.eps} $$

  \caption{ Evolution of the NS in the binary 
systems of figure 1 in the gravitational mass-radius plane.
The mass is in solar masses and the radius is in kilometers. The numbered dots 
indicate evolutionary stages with system age since the start of the accretion  
in units of $10^8$ yr (external numbers) and the spin period in milliseconds 
(internal, bold numbers). These numbers refer to simulations in the 
absence of gravitational waves emission.
The thin dot dashed line is the maximum rotation limit, the dashed thin line 
indicates the stable nonrotating configuration and the thin solid line 
indicates the stability limit to gravitational collapse. The thick solid 
line marks the evolution during the accretion phase in absence of gravitational waves emission,
 the thick dot dot dashed line marks the evolution during the accretion phase,  with gravitational waves driven
by the r-modes instability cycle, 
while the thick dashed line marks the pulsar phase. In case a) the pulsar evolution is shown until the 
star reaches a spin period comparable to that of the fastest pulsar 
observed, PSR~B1937+21, that is 1.56 ms and then slowly brakes to 16 ms in $\sim 8 \times 10^{11}$ yrs.
In case b) pulsar evolution, after a relatively short radio-pulsar phase lasting $5 \times 10^7$ yrs, 
brings the star to  gravitational collapse as it crosses the stability limit. 
In case c) there is no pulsar phase since the accretion of matter brings the 
star directly to gravitational collapse.
Case d) features the same companion star as case c), but the NS is governed 
by the ultrastiff EOS N. The evolution is similar to case a).
Little jumps along the evolution curve are due to the resolution of the grid of 
relativistic values we used.}
\label{fig2}
\end{figure*}

In these evolutions the mass trasfer rate is smaller than the Eddington limit.
 Therefore it is reasonable to assume that the mass transfer is conservative.  
Moreover, the matter flow from the companion is not subject to large fluctuations.
 The NS mass and spin 
evolution in more complicated cases, like the ones observed in X-ray 
transients (see Campana et al. 1998 for a review) or the ones discussed by
Burderi et al. (2001), are not considered here and will be discussed in a 
following paper.
We considered both the case in which the accretion process does not excitate the r-mode instability, 
and the case in which the spin evolution is
influenced by the gravitational waves emission.

In figure 3 we plot the evolution of these systems in the gravitational 
mass -- radius plane. The area of this plane that a stable NS equilibrium 
configuration can span is limited from below by the sequence of stable 
non-rotating equilibrium configurations, which are the ones with 
minimum radius for a given gravitational mass, and  from above by the 
mass-shedding sequence -- that is the sequence of equilibrium configurations 
for which the NS angular velocity equals the Keplerian angular velocity at 
the NS surface radius.
These two stability lines are connected by a third curve, the secular 
stability curve, which limits from the right the region in which 
equilibrium configurations are stable against gravitational collapse.

The system evolution  without gravitational waves emission are 
indicated by the thick solid and dashed lines in figure 3, where the numbered 
dots give the corresponding age and spin period.  We see that the first 
phase of accretion (thick solid lines) is similar in all cases: a rapid spin up brings 
the star to the mass-shedding limit. Thus  
the star remains on the mass shedding sequence in the M--R diagram; it will 
detach from it only when the accretion stops and the NS switches on as a 
radio pulsar (thick dashed lines).

When we consider the spin-down due to r-modes excitation, all evolutionary tracks (dot-dot-dashed lines in 
figure 3)
are even more similar: after a brief spin up phase, the NS attains the critical spin period of 2 ms. 
After that, the spin-up spin- down cycle begins and it ends either in a direct collapse to a black hole or in a radio pulsar phase.
In both hypotheses, the amount of mass accreted 
before the pulsar lights up will determine the fate of the system.

\begin{figure*}
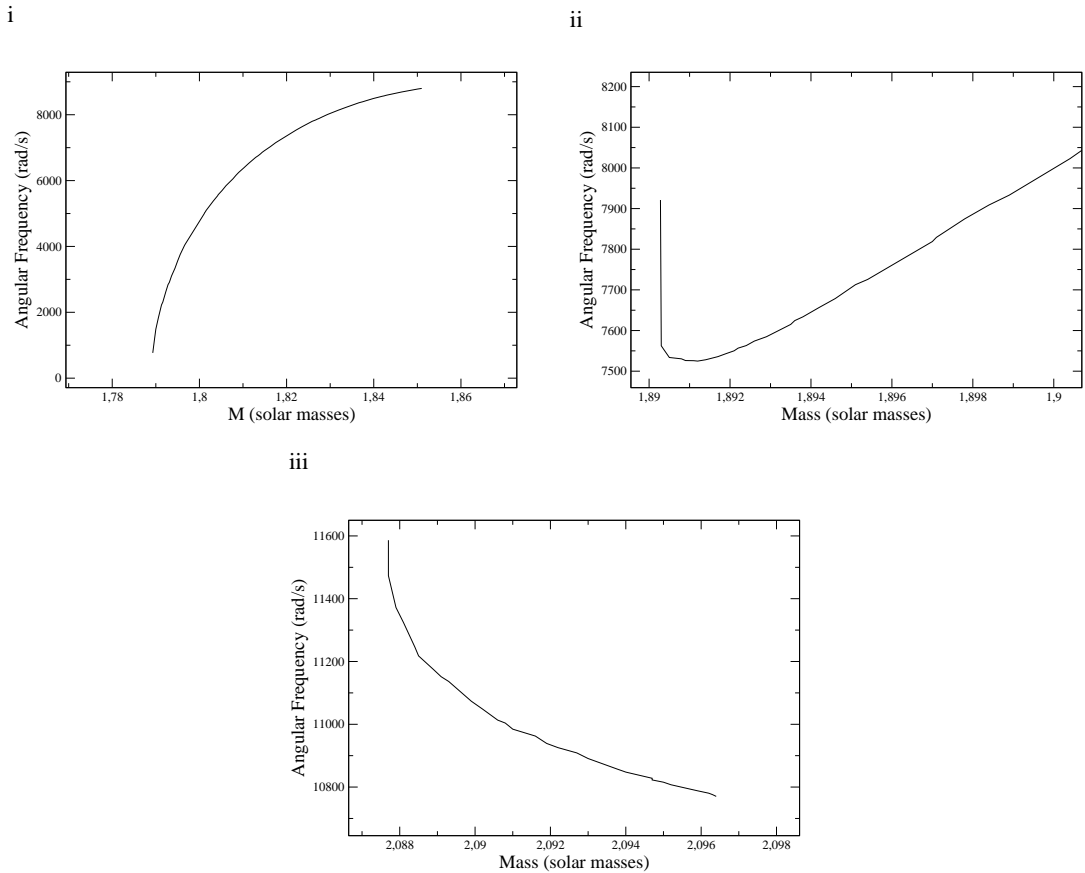

  \centering
$$ \epsfig{figure=fig4a.eps} \qquad \epsfig{figure=fig4b.eps}$$
$$ \epsfig{figure=fig4c.eps}$$

  \caption{ Different evolutionary tracks for a pulsar in the 
$M_G - \omega$ plane depending on its baryonic mass. 
i) NS with a baryonic mass equal to that at the end of accretion in the evolutionary case a.
  The baryonic mass 
is low enough so that the rotating star has a stable nonrotating counterpart 
and the derivative $({\p \omega}/{\p M_G})_{M_B}$ is always positive. 
ii) NS in the supramassive sequence. In this case, 
$({\p \omega}/{\p M_G})_{M_B} \rightarrow - \infty$ 
near the onset of the secular instability, as noticed by various authors, 
but is still positive away from it. The spin up phase for a pulsar of this 
mass, assuming a magnetic field of $\sim 10^8~G$, lasts $\sim 10^6 \mathrm{yr}$. 
iii) NS in the extremely supramassive regime, with a baryonic mass equal to that at the end of accretion
in evolutionary case b. In this case, the mass is so high that 
$({\p \omega}/{\p M_G})_{M_B} < 0$ for any stable configuration. 
The spin up phase, as can be seen from our simulations, lasts 
$\sim 5 \times 10^7~\mathrm{yr}$.
}
\label{fig3}
\end{figure*}

In figure 4 we show the possible behaviors of NS angular velocity as a 
function of gravitational mass for given values of baryonic mass. 
From equation (\ref{eq:evpul}) we see that during the pulsar phase
the spin evolution of the star will only depend on the baryonic mass 
of the star and on the sign of the derivative $({\p \omega}/{\p M_G})_{M_B}$.
 If there is a non-rotating stable configuration for
the given baryonic mass, then $({\p \omega}/{\p M_G})_{M_B} > 0$ always 
and the pulsar will spin down until it stops. If such a configuration does
not exists for the given baryonic mass, two different evolutionary tracks 
are possible: either the pulsar spins down until it comes close to
the stability limit to gravitational collapse, where it spins up rapidly (in fact 
$({\p \omega}/{\p M_G})_{M_B} \rightarrow -\infty$ at the instability), or it
spins up until it collapses, with no spin down phase 
(i.e. $({\p \omega}/{\p M_G})_{M_B} < 0$ always). We say that in the 
former case the pulsar lights up in the supramassive regime, while in the 
latter it lights up in the extremely supramassive regime. 
The spin-up phase, that occurs in correspondence to a loss of rotational 
energy, is caused by the fact that, near the onset of the instability and in 
the extremely supramassive regime, a little loss of energy (gravitational 
mass) corresponds to a rapid decrease of the NS radius. Since the
momentum of inertia of the NS strongly depends on NS radius R and on its compactness 
$G M_G / R c^2$ ($I \sim 0.21 M_G R^2 / (1 - 2 G M_G / R c^2)$, see Ravenhall 
\& Pethick 1994), in these regimes the loss of rotational energy is achieved 
by means of a reduction of the momentum of inertia rather than via a spin down of the NS, that, 
instead, will spin up to partially compensate the reduction of the momentum of inertia.
 
In case a, in the absence of r-mode excitation, the pulsar lights up, with a period of 0.71 ms, when the star has 
a gravitational mass larger than the maximum non-rotating mass i.e. $M_\mathrm{stat}=1.803 M_\odot$ (see figure 4i). 
As the pulsar loses energy due to dipole radiation according to equation 
(\ref{eq:evpul}), it leaves the mass-shedding sequence and returns in the 
normal gravitational mass range, below $M_\mathrm{stat}$. Thus the pulsar will slow down (see figure 3a), without 
collapsing, until it almost stops on a very long time-scale 
(it reaches $16~\mathrm{ms}$ in $\sim 8 \times 10^{11}$ yr).
 If the r-modes are excited during the accretion process, the star lights up as a pulsar 
with the same baryonic mass (since the same amount of mass has been accreted), but with 
a different gravitational mass, 1.792 M$_\odot$ instead of 1.805 M$_\odot$: the larger rotational
period of the newborn pulsar, 3.31 ms, with respect to the period of 0.71 ms of the evolution in absence of gravitational waves,
 means that there is not enough energy
to increase the gravitational mass above the maximum non-rotating mass.
 In this case, the pulsar slows down on a similar long timescale,
reaching  $16~\mathrm{ms}$ in $\sim 7 \times 10^{11}$ yr.

In case 2, if r-modes are not excited the star becomes a pulsar when it is in the extreme 
supramassive regime (see figure 3b) and thus equations (\ref{eq:evpul}) 
imply that the NS is spinning up rather than spinning down. 
As it loses energy, however, the star begins to shrink and heads towards the 
secular instability limit which brings the NS to the gravitational collapse. 
Its radio pulsar phase, characterized by periods well below 1 ms 
(its initial spin period is 0.581 ms, its period when it crosses the stability 
line is 0.542 ms) and by the unusual sign of the period derivative, lasts 
$\sim 5 \times 10^7 {\rm yr}$. Such a short lifetime, 
if compared with typical lifetimes for spinning down pulsars, is due to 
the positive feedback we get in this case for the second of equations 
(\ref{eq:evpul}): gravitational mass loss causes a spin up, which in turn will make the 
term on the right of this equation bigger, causing a further mass loss.
If the r-modes are excited no such behavior arises, since spin of the NS is kept above 2 ms,
the maximum sustainable mass is smaller than that for a NS rotating at mass shedding
(the maximum sutanaible gravitational mass for a NS  spinning at 
a period of 2 ms is $1.817 M_\odot$). Thus the star collapses to a black hole 
 $\sim 5 \times 10^8$ yr before the end of the accretion process.

In case c, the star never lights up as a pulsar, as too much matter is accreted 
on the star and the maximum mass limit is exceeded. Thus the NS will directly
collapse to a black hole  (see figure 3c).

All this behaviors are strongly dependent on the EOS adopted to describe the 
ultradense matter. As a comparison we studied the evolution of a system 
consisting of the same companion star as in case c
and a $1.4 M_{\odot}$ NS with an ultra-stiff EOS (EOS N by Walecka \& Serot, 
see Arnett and Bowers 1977), which has a maximum non-rotating mass of 
$2.634 M_{\odot}$. The result of this evolution  different, 
as we show in figure 3d; in this case (hereafter case d) no spinning up pulsar 
shows up, nor any accretion induced collapse happens. Instead, we end up with 
a spinning down submillisecond pulsar with a very long lifetime, 
comparable with the one we obtained in the first case. If the r-modes are 
excited (see again figure 3d) the minimum attanaible period is limited 
to 2 ms, and the pulsar lights up with a period of 2.27 ms.
In a following paper, we will discuss in full detail the effects of the EOS of the
ultradense matter on the evolution of LMXBs.

\section{Conclusions}

We have shown that, depending on the characteristics of the system, 
especially on the amount of mass accreted on to the NS,  on the EOS adopted to describe the
Ns matter, and on the excitability of the r-modes, 
LMXBs can have quite different fates: they can light up as a spinning down
 radio pulsars,  they can directly 
collapse to a black hole during the accretion phase or if, at the end of the accretion phase,
 the NS is left in the extreme supramassive regime, it will light up as an 
exotic, spinning up submillisecond radio pulsar with a relatively short lifetime.

It is then evident that, in the hypothesis of a conservative mass transfer
from the companion  onto a low--magnetized NS and in absence of r-modes excitation, 
the accretion process, if the amount of mass accreted is not enough to collapse into a black hole,
will end with a very fast spinning object,
as it has been suggested before (Cook, Shapiro, \& Teukolsky 1994a).
If the NS becomes then detectable as a radio pulsar, it will have a spin period well below one millisecond.
In fact our simulations show (cases a and d) that we obtain submillisecond pulsars with long 
lifetimes (in the former case  the pulsar lifetime before reaching 
a period as long as that one of the fastest millisecond pulsar known 
to date, PSR~B1937+21, is $\sim 3 \times 10^{9}~{\rm yr}$, while in the 
latter it becomes $\sim 5 \times 10^{9}~{\rm yr}$). 

If r-modes are excited by accretion, pulsars are constrained to spin slower than a critical frequency, and this
could  explain why no NS spinning at submillisecond periods has been observed to date. However, 
in this situation any binary system in which enough mass is transferred from the companion to the
NS will collapse to a black hole, without lighting up as a pulsar. Thus pulsar formation could
be much less favoured than in other cases.

It is likely that millisecond pulsar systems like the ones observed to 
date (i.e. systems with $P> 1~\mathrm{ms}$) 
 originate from different binary evolution scenarios,  in which some critical mechanism 
has prevented the accretion process to continue until a mass as large as 
a significant fraction of a solar mass has been transferred. It is probable that the magnetic field of the NS has 
 values much higher than the value we chose (at least at the beginning of the accretion), so that the inner edge of 
the disc could be outside the corotation radius for at least part of the evolution and 
magnetic torques could play an important role in the spin evolution of the NS. Moreover,
systems in which the mass transfer rate has large fluctuations will 
light up as pulsars before the end of the accretion process, losing a 
large amount of mass in a so-called radio ejection phase as proposed by Burderi et al (2001). We will
investigate the evolution of such systems in a future paper.

Although there are 
selection effects that could have prevented the discovery of 
submillisecond pulsars (Burderi et al 2001), if a self-limiting mechanism like the r-mode instability does not operate,
 submillisecond radio pulsars should exist and should be detectable in the future. 
On the other hand we have shown (case 3) that if the NS matter is governed
by a moderately stiff EOS like  FPS (i.e. with a maximum non--rotating mass of $\le 1.9 M_\odot$), 
the mass transfer can end in an accretion induced collapse to a black hole if as much as $1 M_\odot$ is
accreted. Although not much in known on the range of progenitor masses fot today's population of LMXB in the Galaxy,
a recent study by Pfahl, Rappaport and Podsialowski (2003) argue that a great fraction of observed LMXBs may have
descended from intermediate mass X-ray binaries, that is form systems with initial donor mass $\ga 1.5 M_\odot$. If this is true,
and if we assume that in such a system
we have equal probability that the companion transfers any amount of mass between $0.5~M_\odot$ and $ 2~M_\odot$ 
on to a $1.4~M_\odot$ NS, for NS governed by EOS N we have a $100\%$ probability of obtaining a  submillisecond pulsar, while for NS governed by EOS FPS
 we have only a $20\%$ probability of obtaining a spinning down millisecond pulsar,
a $9\%$ probability of the formation of a spinning up submillisecond pulsar (doomed to gravitational collapse)
 and a $71\%$ probability of  direct collapse into a black hole during the accretion phase.
Therefore if the EOS governing NS matter is soft, a conservative mass transfer 
is more likely to end with a direct accretion-induced collapse
to a black hole than with the formation of a submillisecond radio pulsar, and thus submillisecond pulsars 
could be hard to detect because of their low formation probability. 

On the other hand, if the r-modes are excited, the spin period will remain 
well above one millisecond for all of the evolution. However, making the same assumptions as before,
 the probability of forming a pulsar drops to $10 \%$ for EOS FPS, while for EOS N
we still have a very high probability, $ \sim 90 \%$. Therefore this scenario, in which the mass transfer is 
conservative but the spin frequency is limited by the emission of gravitational waves, 
implies that the EOS is stiff in order to have an high probability of formation of millisecond pulsars.

Teherefore we should predict that the EOS of NSs is very stiff in order to explain the observational evidence (MSP are formed), 
if gravitational waves are indeed emitted due to r-modes excitation, while we should predict that
 the EOS of  NSs is soft if they are not emitted, so that submillisecond pulsars are very uncommon, as the observations seem to indicate.
 If future observations will allow to constrain the stiffness of the NS EOS on an observational basis,
this will give an indication on wherther r-modes are indeed excited in LMXBs or not.

We have shown that, if r-modes are not excited in LMXBs, the accretion process can leave us with an
 extremely supramassive NS, that will spin up during all of its life as a radio pulsar (case b).
It is evident that, being the critical baryonic mass  for getting 
to the extremely supramassive regime $M_{\mathrm{crit}}$ an EOS-dependent feature, in 
principle the observation an accelerating (or braking) submillisecond 
pulsar can allow to exclude several EOS on an observational basis. 

\begin{figure}
  \centering
  \epsfig{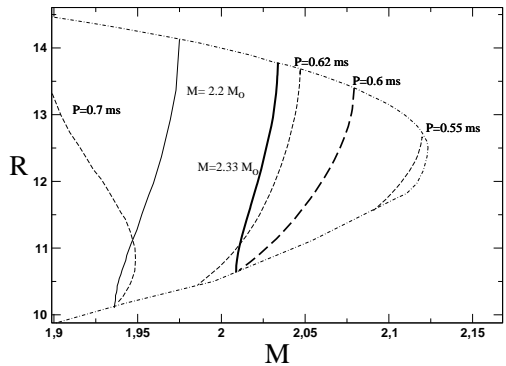}
  \caption{
Sequences of NSs with the same spin period are plotted as dashed lines in the  gravitational mass-radius plane, while sequences of NSs
with the same baryonic massare plotted as solid lines. We consider NSs governed by EOS FPS.
 The dashed--dotted line limits stable configurations. We plot with a thick solid line the sequence of Ns with baryonic mass equal to the critical mass
, $M_{\mathrm{crit}}=2.33 M_\odot$. Any star with $M_B \ge M_{\mathrm{crit}}$ is extremely supramassive, i.e. it spins up under magnetodipole radiation.
It is interesting to note that while sequences of constant baryonic mass always have the same shape 
in the gravitational mass-radius plane, bending from left to right with increasing radii, sequences of constant spin frequency have completely different topologies
below the critical mass and above it (see for example the sequence with $P= 0.7~\mathrm{ms}$ and the one with $P=0.6~\mathrm{ms}$).
During the radio pulsar phase the star moves along a sequence of constant baryonic mass, decreasing its gravitational mass. It moves therefore from top right 
of the figure to the bottom left. This implies that the pulsar spins down as long as constant spin frequency sequences bend from top left to bottom right 
(as the one with  $P= 0.7~\mathrm{ms}$), and that it spins up if the  constant spin frequency sequences it crosses bend from
bottom left to top right in the plane (as the one with $P= 0.6~\mathrm{ms}$ does).
As shown in the figure, any stable NS attaining a period $P \le P_{\mathrm{crit}}= 0.6~\mathrm{ms}$ (i.e. any 
star who lies on the right of the thick dashed line) has $M_B \ge M_{\mathrm{crit}}$. Thus any NS
governed by EOS FPS attaining a period $\le 0.6~\mathrm{ms}$ will spin up once it becomes a pulsar.
}\label{fig5}  
\end{figure}

To clarify how such an effect can help to constrain the EOS of ultradense matter
we need to introduce the new concept of a \textit{critical spin period} $P_{\mathrm{crit}}$ that,
together  with the minimum period ($P_{\mathrm{min}}=2 \pi / \omega_{\mathrm{max}}$), is peculiar to each EOS.
 $P_{\mathrm{crit}}$ is the period 
below which the EOS allows only extremely supramassive stable configurations.
In figure 5 we show sequences of equilibrium configurations
with constant spin period, together with the critical baryonic mass sequence. It is evident 
from the figure that $P_{\mathrm{crit}}$  is  equal to the minimum allowed period
to avoid gravitational collapse if the star has $M_B=M_{\mathrm{crit}}.$
In fact, any constant period sequence  $P < P_{\mathrm{crit}}$
will only include stars of baryonic mass greater than the critical one.
 Thus any NS with  $P < P_{\mathrm{crit}}$  will
accelerate as a consequence of energy loss due to magnetic dipole radiation. Being $P_{\mathrm{crit}}$ EOS-dependent, 
the detection of a submillisecond radio pulsar and the determination of the sign of its period 
derivative will allow to effectively constrain the equation of state 
governing ultradense matter.

Thus the detection of a submillisecond radio pulsar can impose two constraints on the EOS of the NS:
\begin{enumerate}
\item the spin period must be larger than the minimum allowed period, i.e. the spin period of the maximum
  rotation configuration,  $P_\mathrm{min}$. 
\item if the period is shorter than $P_{\mathrm{crit}}$, the radio pulsar must spin up rather than spin down.
\end{enumerate}
 Both $P_\mathrm{min}$ and $P_{\mathrm{crit}}$ are EOS dependent and are longer for stiffer EOSs.
In fact, the detection of a submillisecond radio pulsar with spin period $P_\mathrm{obs}$ undergoing a
spin up will rule out all the stiff EOSs with  
 $P_\mathrm{min}> P_\mathrm{obs}$. On the other hand the detection of a spinning 
down submillisecond radio pulsar,  with spin period $P_\mathrm{obs}$, 
will allow us to rule out all the stiff EOSs with 
$P_\mathrm{crit}> P_\mathrm{obs}$, because they cannot explain a spinning down radio pulsar with such a short spin period.
In this case the limit is more stringent because $P_\mathrm{crit}> P_\mathrm{min}$!
As an example, suppose that  a spinning down radio pulsar with 
a period of $0.713$ ms  (like the one we obtain in case a) will be detected: 
this will allow us to rule out EOS N, since although the minimum period for 
this EOS is $0.69$ ms, any radio pulsar governed by EOS N with such a low spin 
period will spin up, being for EOS N $M_{\mathrm{crit}}=3.63 M_\odot$ and $P_{\mathrm{crit}}=0.74 ~\mathrm{ms}$.  

In summary, in this paper we presented the first results obtained with a new code that allows to study in  details
the binary system evolution and the spin evolution of the NS, on the basis of fully relativistic calculations.
We used this code to study the evolution of systems with conservative mass transfers and confirmed that the large amount of matter 
that is transferred on to the NS will spin it up to periods well below one millisecond, unless the emission of gravitational waves 
dissipates the excess of angular momentum.
 However in this last case the amount of mass acreted onto the NS is easily big enough to cause a direct collapse to a black hole.
 Therefore we concluded that presumably the recycled systems which give origin to the MSP observed to date should have origin from 
systems with a highly nonconservative mass transfer.
We showed that if the EOS of ultradense matter is not very stiff the direct collapse to a black hole is very likely to happen even if the r-modes
are not excited. This could explain the lack of any observation of submillisecond radio pulsars even without 
invoking gravitational waves emission.
As a last remark, since we showed that there is the possibility of obtaining from binary evolution some unusual, accelerating 
submillisecond radio pulsars, we introduced the new concept of the critical spin period $P_{crit}$, peculiar to each 
EOS, that can allow to effectively constrain 
 the EOS of NS matter if a radio pulsar with a period below one millisecond will be observed in the future.  
\section*{Acknowledgements}
 We thank the anonymous referee for useful discussions that helped us improve this paper. This work was partially supported by the 
Ministero dell'Istruzione, dell'Universit\`a e della Ricerca (MIUR) with PRIN 2001 and by the Universit\`a di Palermo.

\end{document}